# Tachyon motion in a black hole gravitational field

V. M. Lipunov

ABSTRACT

The motion of superluminal particles in the gravitational field of a non-rotating black hole is analyzed. The relativistic Hamilton-Jacobi equation is solved for particles with imaginary rest mass. It is shown that there are no stable circular orbits and generally no finite motions for tachyons in the Schwarzschild metric and that all unstable circular tachyon orbits lie in a region extending from the gravitational radius to 1.5 times that radius. The particles with speeds exceeding the speed of light are noticed to be able to escape from the space limited by the gravitational radius. The results also indicate that low-energy tachyons near a black hole may acquire higher energies and that this in turn may lead to observable effects.

Over the past 10 – 15 years a number of papers have been published [1, 3, 5, 6] about objects moving at superluminal velocities. The usual arguments against the existence of such objects proved to be not so indisputable as they seemed.

Abandoning the requirement for the rest mass, proper lengths, and proper time to be real quantities allows suggesting the existence of superluminal particles. Note that in this case all observables – energy, momentum, and coordinates – remain real. Such particles, which are called tachyons by G. Feinberg, have a number of unusual properties. The rest mass of a tachyon is usually written in the form:

$$m_o = i\, m_*,  \qquad (1)$$

where $m_*$ is a positive quantity. The energy of a tachyon moving at velocity $v>c$ is the given by the following formula:

$$E = \frac{m_* c^2}{\sqrt{v^2/c^2 - 1}}. \tag{2}$$

The energy of a tachyon decreases with increasing velocity and becomes zero in the limit of infinite velocity. The state of tachyons with zero energy is usually called transcendent. In the transcendent state the tachyon has a finite momentum equal to $m_* c$.

Numerous experiments aimed at detecting tachyons failed to produce positive results. If such particles exist, this may imply that they interact only weakly with ordinary particles. However, tachyons may have been generated intensively during early stages of the cosmological expansion (the Planck epoch) leaving a relic background of tachyons transcendent with respect to the comoving frame. The velocity of such tachyons relative to the laboratory frame can be estimated as:

$$v \simeq c^2/w,$$

where $w \approx 200$ km/s is the velocity of the Earth with respect to cosmic microwave background. The velocity of background tachyons should therefore be $v \approx 4.5 \times 10^8$ km/s. Hence, in terrestrial laboratory experiments, we are dealing with tachyons that have energies $\ll m_* c^2$.

Without making any assumptions concerning the magnetic or electric properties of tachyons we can conclude that strong gravitational fields are needed to increase substantially the energy of background tachyons. In this respect, ideal objects for pumping energy to tachyons are black holes in whose neighborhood interaction of tachyons with ordinary particles may produce observable effects. Note that the allowance for the interaction with tachyons in the vicinity of a true singularity maybe especially important.

In this paper we consider the motion of tachyons in the gravitational field of a nonrotating black hole. This problem for ordinary particles was solved before [2].

We do not take into account the gravitational radiation of tachyons, which should have the form of shock waves, and use the Hamilton–Jacoby equation [4] to determine the trajectory:

$$g^{ik} \frac{\partial S}{\partial x^i} \frac{\partial S}{\partial x^k} + m_*^2 c^2 = 0. \tag{3}$$

Here $g^{ik}$ are the components of metric tensor; $x^i$ and $x^k$ are coordinates, and $S$ can be written as:

$$S = -\varepsilon_0 t + M \varphi + S_r(r), \tag{4}$$

where $\varepsilon_0$ and $M$ are the energy and angular momentum of the particle at infinity. We consider the motion in the $\theta = \pi/2$ plane.

Let us assume that tachyons obey the same equations of motion as ordinary particles with the underlying assumption of the equality of the inertial and gravitational mass.

In the Schwarzschild metric, equation (3) acquires the following form:

$$\left(1 - \frac{r_g}{r}\right)^{-1} \left(\frac{\partial S}{c \partial t}\right)^2 - \left(1 - \frac{r_g}{r}\right) \left(\frac{\partial S}{\partial r}\right)^2 - \frac{1}{r^2} \left(\frac{\partial S}{\partial \varphi}\right)^2 + m_*^2 c^2 = 0, \tag{5}$$

where $r_g = 2GM/c^2$ is the gravitational radius of the central body. We then derive from equations

$$\frac{\partial S}{\partial \varepsilon_0} = const \quad \text{and} \quad \frac{\partial S}{\partial M} = const$$

the following dependences $t(r)$ and $\varphi(r)$:

$$t = \frac{\varepsilon_0}{c^2} \int \frac{1}{1 - \frac{r_g}{r}} \left[\frac{\varepsilon_0^2}{c^2} - \left(\frac{M^2}{r^2} - m_*^2 c^2\right)\left(1 - \frac{r_g}{r}\right)\right]^{-1/2} dr, \tag{6}$$

$$\varphi = \int \frac{M}{r^2} \left[\frac{\varepsilon_0^2}{c^2} - \left(\frac{M^2}{r^2} - m_*^2 c^2\right)\left(1 - \frac{r_g}{r}\right)\right]^{-1/2} dr. \tag{7}$$

We now introduce the following notation

$$E_0 = \varepsilon_0 / m_* c^2, \quad \mu = M / m_* c r_g, \quad \rho = r / r_g$$

for the sake of convenience. The radial component of tachyon velocity measured by a stationary observer at each point of the tachyon trajectory can be determined from equation (6):

$$v_r = c \frac{\left(E_0^2 + 1 - \frac{1}{\rho} - \frac{\mu^2}{\rho^2} + \frac{\mu^2}{\rho^3}\right)^{1/2}}{E_0}. \tag{8}$$

If the tachyon moves radially (μ=0), its velocity decreases as it approaches the event horizon, and at the horizon it becomes equal to the velocity of light like in the case of ordinary particles. Consequently, the tachyons with low energy at infinity may acquire large energy in the vicinity of the black hole. This energy may be entirely or partially emitted in the form of gravitational waves. For an observer at infinity, all processes freeze in the vicinity of the event horizon and the coordinate velocity of the tachyon $dr/dt \rightarrow 0$.

Let us now introduce the effective potential energy:

$$U(\rho) = \left[ -\left(1 - \frac{1}{\rho}\right)\left(1 - \frac{\mu}{\rho}\right)\left(1 + \frac{\mu}{\rho}\right) \right]^{1/2}. \tag{9}$$

Equation (8) can then be rewritten in the form:

$$v_r = \frac{1}{E_0}\left[E_0^2 - U^2(\rho)\right]^{1/2}. \tag{10}$$

The radial velocity becomes zero at the turning point:

$$E_0 - U(\rho_t) = 0. \tag{11}$$

Figure 1 shows qualitatively the dependence of potential energy on distance at different time instants. The potential curve has no minima in the $1 < \rho < \infty$ domain, and therefore tachyons have no stable circular orbits. The maximum of the $U(\rho)$ curve corresponds to an unstable circular orbit and is determined by the following equation:

$$U'(\rho)\big|_{\rho = \rho_m} = 0. \tag{12}$$

We now solve the last equation for the coordinate of the maximum to obtain:

$$\rho_m = \mu^2\left(-1 + \sqrt{3/\mu^2 + 1}\right). \tag{13}$$

Figure 2 shows the dependence of $\rho_m$ on angular momentum $\mu$ (curve 1) and the corresponding dependence for ordinary particles (curve 2). Circular orbits for tachyons begin at the distance of $(3/2)\, r_g$, i.e., at the distance of the innermost unstable orbit of ordinary particles. Note that the innermost unstable circular orbit for a tachyon is located at the gravitational radius and corresponds to a tachyon that is transcendent at infinity and has angular momentum $M = m_* \, c \, r_g$. Transcendent tachyons with smaller angular momenta go under the gravitational radius and therefore the capture cross section σ for transcendent tachyons is given by the following simple formula:

$$\sigma = \pi \, r_g^2,$$

i.e., transcendent tachyons do not "feel" gravitation.

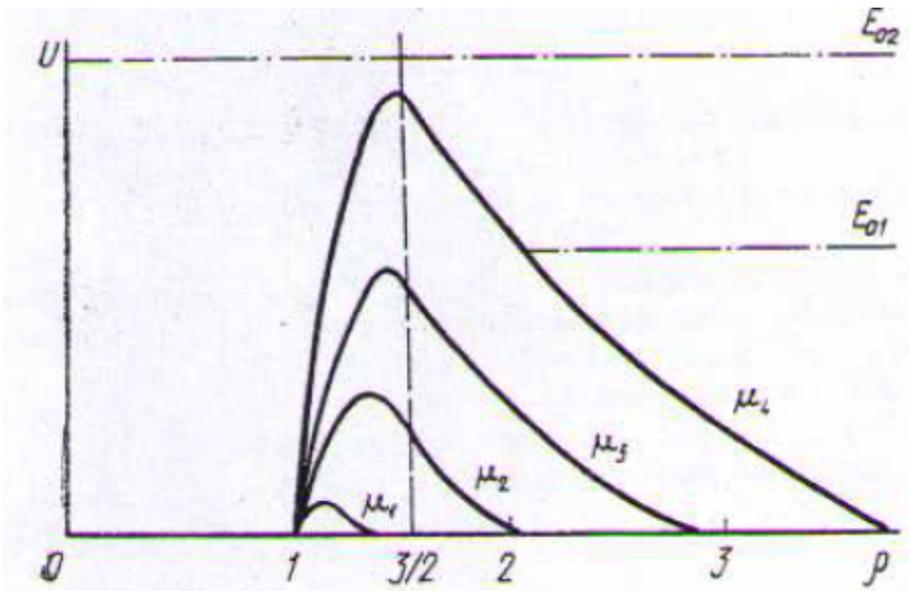

Fig.1. Dependence of the effective potential energy on distance for different angular momenta ($\mu_1 < \mu_2 < \mu_3 < \mu_4$). The tachyon with energy $E_{02}$ moves under the gravitational radius without encountering a turning point.

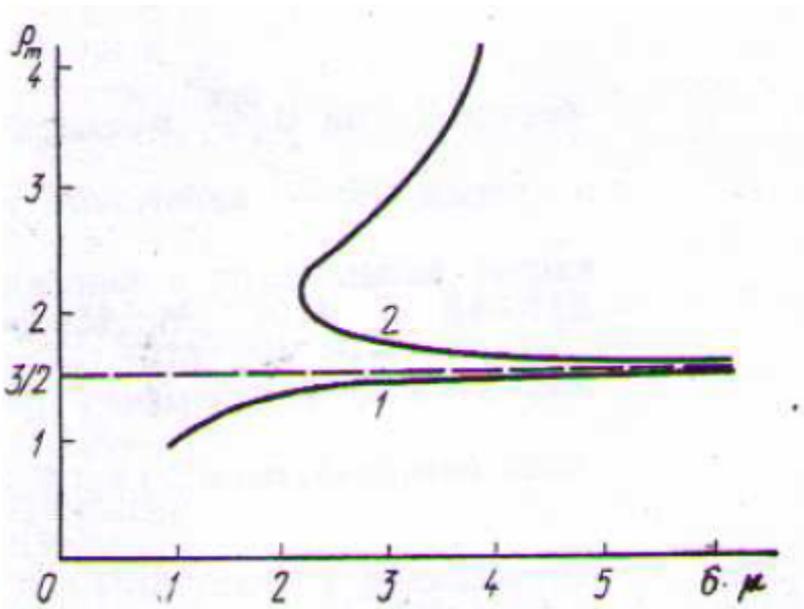

Fig. 2. Dependence of the radius of the circular orbit of a tachyon on its angular momentum (1) and the corresponding dependence for ordinary particles (2) [4].

It is interesting that tachyons may emerge from under the event horizon. An observer located in the vicinity of a black hole could "see" tachyons appearing that move at the speed of light. This is easy to verify by considering the propagation of light signals in the Lemaitre metric. The trajectories of tachyons lie outside the light cone and therefore these particles may move away from the center even in a collapsing T region.

I am grateful to Z.P.Shengait for useful discussions.

Main Astronomical Observatory of the Academy of Sciences of the Ukrainian SSR.